**Rotation-assisted wet-spinning of UV-cured gelatin fibres and nonwovens**

Jessica Rickman,[1,2] Giuseppe Tronci,[1,2] He Liang,[1,2] and Stephen J. Russell[1]

[1]Clothworkers' Centre for Textile Materials Innovation for Healthcare (CCTMIH), School of Design, University of Leeds, Leeds, LS2 9JT, UK

[2]Department of Oral Biology, School of Dentistry, University of Leeds, Wellcome Trust Brenner Building, St James' University Hospital, Leeds, LS9, 7TF, UK

Correspondence: sd14jr@leeds.ac.uk

**Abstract**

Photoinduced network formation is an attractive strategy for designing water-insoluble gelatin fibres as medical device building blocks and for enabling late-stage property customisation. However, mechanically-competent, long-lasting filaments are still hard to realise with current photoactive, e.g. methacrylated, gelatin systems due to inherent spinning instability and restricted coagulation capability. To explore this challenge, we present a multiscale approach combining the synthesis of 4-vinylbenzyl chloride (4VBC)-functionalised gelatin (Gel-4VBC) with a voltage-free spinning and UV-curing process so that biopolymer networks in the form of either individual fibres or nonwovens could be successfully manufactured. In comparison to state-of-the-art methacrylated gelatin, the mechanical properties of UV-cured Gel-4VBC fibres were readily modulated by adjustment of coagulation conditions, so that an ultimate tensile strength and strain at break of 25±4–74±3 MPa and 1.7±0.3–8.6±0.5 % were measured, respectively. The sequential functionalisation /spinning route proved to be highly scalable, so that one step spunlaid formation of fibroblast-friendly nonwoven fabrics was successfully demonstrated with wet spun Gel-4VBC fibres. The presented approach could be exploited to generate a library of gelatin building blocks tuneable from the molecular to the macroscopic level to deliver computer-controlled extrusion of fibres and nonwovens according to defined clinical applications.

# 1. Introduction

As derived from the partial denaturation of collagen, gelatin has significant potential as a building block for economically attractive manufacture of medical devices, due to its biocompatibility, wide availability and low cost. As well as regenerative devices, the use of gelatin encompasses wound dressings, drug delivery, vascular prostheses, orthopaedic membranes and skin substitutes (1-12).

Gelatin is a polypeptide formed by the denaturation of collagen, often thermally or chemically, wherein the hydrogen bonds stabilising the triple helices are disrupted. This process leads to triple helix disentanglement into separate randomly organised polypeptide chains. Gelatin therefore has the same primary amino acidic structure as collagen, which makes gelatin biodegradable and compatible with the physiological environment due to the presence of collagenase-cleavable and cell-binding peptide sequences, respectively (13, 14).

Spinning of gelatin fibres for use in medical devices has been reported via dry spinning, gel spinning, electrospinning and wet spinning (10, 15-21), yielding varied fibre properties. Electrospinning or wet spinning are typically used to produce gelatin fibres (17, 18, 20, 22). Electrospinning is a one-step process employing electrostatic voltage to achieve a mesh of submicron fibres. In light of the submicron fibre diameter, electrospun fibres are advantageous for mimicking fibres found in native tissue more closely and for realising 2D fabrics (overall thickness < 1 mm) with increased total surface, wet-spinning is voltage-free and yields individual, dimensionally-controllable fibres that can be subsequently converted in to a wider range of textile formats, including nonwovens as well as knitted, woven, braided and hybrid structures. Compared to electrospinning, wet-spun fibres assembled into nonwovens allow for increased customisation of liquid absorbency, porosity, mechanical properties, cell homing capability, and macroscopic shape (23-26). Other major advantages of wet spinning are the ability to more readily control and predict molecular alignment and the fibre diameter distribution using fibre drawing, dope concentration alteration and spinneret size compared to electrospinning (27-29). Consequently, wet spinning provides a more consistent, yet highly versatile route to industrial manufacture of fibre-based medical devices compared to a platform based on electrospinning.

The functionality of gelatin fibres in medical devices is impeded by gelatin's uncontrollable swelling at room temperature in aqueous environments, thermo-reversible dissolution at increased temperature and poor thermo-mechanical properties in physiological conditions. Taking inspiration from nature, covalent functionalisation and crosslinking of gelatin chains have been pursued to restore covalent crosslinks present in collagen fibres in vivo and ensure retention of fibrous architecture following material contact with water. Various approaches have therefore been reported in the literature based on either bifunctional reagents, e.g. diisocyanates, (30) and aldehydes (31), or carbodiimide-induced intramolecular crosslinking (18, 32).

Recent work on collagen-derived polypeptides has focused on the synthesis of covalently crosslinked networks as building blocks of water-insoluble wet-spun gelatin fibres, whose wet-state tensile properties were found to be directly related to the molecular weight of the fibre-forming polypeptide (20). Photoinduced covalent networks have also been synthesised with either gelatin random coils or collagen triple helices (19, 33). Once grafted onto the amino acid residues of the polypeptide, covalently-coupled photoactive moieties can induce the formation of covalent crosslinks when exposed to a photoinitiator and a specific wavelength of light. This photoinduced network formation approach enables *in situ* crosslinking to occur at a much faster rate than typical chemical crosslinking reactions (19, 34-36), combined with cellular tolerability. It also introduces additional precipitation steps to ensure that any cytotoxic, unreacted crosslinker has been removed from the material before processing. One of the most prevalent photoactive compounds is methacrylic anhydride (MA), which has been widely employed to obtain gelatin networks with varied macroscopic material formats (19, 35, 37).

Despite extensive research on MA-functionalised gelatin (Gel-MA) hydrogels, relatively few investigations have been reported on the formation of Gel-MA hydrogel fibres, in part due to the poor mechanical properties and fibre-forming capability of the photoactive fibre precursor (35). Shi *et al.* (19) reported the formation of grooved Gel-MA fibres by microfluidic spinning of the polymer solution into an ethanol bath at -21°C, wherein the surface grooves were reported to encourage cell alignment. Control of phase separation and fibre formation was achieved by spinning mixtures of Gel-MA and alginate into $CaCl_2$, whereby the coagulation bath was selected to induce crosslinking of the alginate chains enclosing the Gel-MA. The Gel-MA was subsequently crosslinked using UV irradiation (38). The sacrificial alginate component was removed using a calcium chelator leaving a hybrid hydrogel filament that could be subsequently assembled into woven and braided structures. Given the multiple steps required, a simpler route for delivering gelatin fibres is ideally required to accelerate scale-up. Additionally, although photocured Gel-MA fibres proved to support cells, the mechanical stability of the material in clinically-relevant medical devices has been shown to be sub-optimal. Comparing the mechanical characteristics of collagen hydrogels crosslinked with different photoactive moieties, Tronci *et al.* found that UV-cured hydrogels made of MA-functionalised collagen to exhibit both lower water uptake and compressive modulus than UV-cured hydrogels made of 4-vinylbenzyl chloride (4VBC)-functionalised collagen (34). Rather than hydrogels, it was therefore of interest to determine if vinylbenzylation is capable of yielding mechanically-competent, wet spun gelatin fibres (39), yet ensuring full cellular tolerability. Accordingly, this study focused on the physical properties of photoactive and UV-cured Gel-MA and Gel-4VBC wet spun fibres and compare their *in-vitro* biocompatibility. To mediate fibre-forming capability and minimise toxicity issues, wet-spinning of collagen-based polypeptides into polyethylene glycol (PEG)-supplemented aqueous solutions has been shown to enhance the thermal and mechanical properties of the fibres as well as cell infiltration and tissue in growth (40-42). PEG can be used as plasticiser and is widely accepted as inert and safe polymer, with various PEG-conjugated drugs been approved by the FDA (43). Therefore, an evaluation of wet-spun Gel-4VBC fibre properties was made using both PEG and aqueous coagulation media.

## 2. Experimental

### *2.1 Materials*

Polysorbate 20, diethyl ether, HCl, and ethanol were purchased from VWR Prolabo Chemicals. Acetone was purchased from Fischer Chemicals. Acetic acid was purchased from Fluka Analytical. Dulbecco's PBS was purchased from Lonza Chemicals. High molecular weight gelatin type A (Bloom strength ~ 270), 4VBC, MA, Triethylamine (TEA), poly($\varepsilon$-caprolactone) (PCL) ($M_n$: 80000 g·mol$^{-1}$), PEG ($M_n$: 8000 g·mol$^{-1}$), PBS (0.1 M, pH 7.4), 2-Hydroxy-4-(2-hydroxyethoxy)-2-methylpropiophenone (I2959), and all other reagents were purchased from Sigma Aldrich. LIVE/DEAD staining kits and cell culture plates were purchased from Thermofisher. Murine L929 fibroblasts were purchased from Sigma Aldrich.

### *2.2 Synthesis of Gel-4VBC and Gel-MA*

Gelatin was dissolved in distilled water at 10% (w/v) at 37°C and neutralised to pH 7.4. Either 25× or 10× molar ratio (with respect to the lysine molar content in gelatin) of TEA and either 4VBC or MA ([monomer]=[TEA]) was added with 1 wt.% of polysorbate 20. Lysine molar content was quantified by 2,4,6-trinitrobenzenesulfonic acid (TNBS) assay and proved to be equal to 3.05×10$^{-4}$ moles·g$^{-1}$ of gelatin. The reaction of gelatin with either 4VBC or MA was carried out at 40 °C for 5 hours, then the reacting mixture was precipitated in a 10× volumetric ratio of pure ethanol and incubated for 24 hours to remove unreacted chemicals.

### *2.3 TNBS assay*

The degree of gelatin functionalisation was quantified using the TNBS colorimetric assay as described by Bubnis *et al* (44). Briefly, 1 ml 4 vol.% NaHCO$_3$ and 1 ml 0.5 vol.% TNBS was added to 11 mg of gelatin sample and shaken at 200 r·min$^{-1}$ at 40 °C for 4 h. 3 ml of 6N HCl solution was added to each sample and 1-hour incubation at 70°C was performed. 5 ml distilled water was added and 20 ml diethyl ether (x3) was used to remove any unreacted TNBS. 5 ml of diethyl ether-extracted solution was collected and diluted with 15 ml H$_2$O. Resulting solution was analysed via UV-Vis spectrophotometry (6315 UV/Visible Spectrophotometer, Jenway, UK) and the absorbance value recorded at 346 nm against a blank sample [39]. These absorbance values were used in conjunction with Equation 1 and 2 to determine the molar content of remaining lysines (per gram of gelatin) and the degree of gelatin functionalisation, respectively:

$$\frac{Moles\ (Lys)}{g\ (gelatin)} = \frac{2 \times Abs \times 0.02}{1.46 \times 10^4 \times p \times m} \qquad \textbf{Equation 1}$$

$$F = \left(1 - \frac{Gel_F}{Gel_N}\right) \times 100 \qquad \textbf{Equation 2}$$

In Equation 1, *1.46×10$^4$* is the molar absorptivity (in M$^{-1}$cm$^{-1}$) of trinitrophenly lysine, *p* is the spectrophotometer cell path length (1 cm), m is the sample weight (0.011 g), 0.02 is the total volume (in litres) of the diluted diethyl ether-extracted solution. In Equation 2, *Gel$_N$* and *Gel$_F$* are the molar lysine content in either native or functionalised gelatin, respectively. Three replicates were performed for each sample.

## *2.4 Viscometry*

Wet spinning solutions were heated to 50 ºC and measured with a modular compact rotational rheometer (MCR 302, Anton Paar, Austria). Measurements were taken at 50 ºC using a 25 mm parallel plate with a 0.3 mm distance between it and the sample plate. An initial 60 s of rotation at a shear rate of 0.1 s$^{-1}$ was performed to stabilise the sample, following which solution viscosity was recorded at shear rates of 0.1-1000 s$^{-1}$ over 5 min.

## *2.5 Wet spinning*

Preliminary wet spinning experiments were carried out manually with a syringe and blunt needle with a 0.8 mm internal diameter. 17.4 mM aqueous solution of acetic acid was used to prepare the wet spinning gelatin (native, Gel-MA, and Gel-4VBC) solutions (15 wt.% gelatin), due to the increased solubility of gelatin at acidic pH, aiming to minimising wet spinning-associated risks of needle blockage (45). The native gelatin and Gel-4VBC wet spinning solutions (held at 50ºC) were extruded into a coagulation bath consisting of 0.1 M PBS supplemented with 20% (w/v) PEG at 20ºC. Following confirmation of the spinnability of functionalised gelatin systems, fibres were extruded using a syringe pump in contact with a rotating coagulant bath at a pump rate of 30 ml·hr$^{-1}$. Spinning parameters for each sample can be seen in Table 1.

Gel-MA fibre controls could be wet spun using a wet spinning solution of 30% (w/v) methacrylated gelatin in 17.4 mM acetic acid. A fibre coagulant consisting of 20% (w/v) NaCl (20 ºC) in distilled water was initially used, in line with previous reports (28, 42, 46). Owing to the rapid solidification and poor fibre formation in the salt-supplemented bath, ethanol was employed as a suitable coagulation bath for wet spinning of Gel-MA. To enhance wet spinning compliance, ethanol was stored at -20 ºC prior to, and placed on an ice bath during, wet spinning (19). Instead of conventional linear take-off, the dopes were extruded into a rotating coagulation bath (rotational circumference = 30 cm) with the needle placed eccentrically in the bath. The rotation speed was selected at 1.6 rpm, whilst the needle tip was 1 cm away from the bath bottom surface. The fibres were incubated in the coagulating bath for 1 hour at 5 ºC to enhance phase separation of gelatin and promote fibre formation. Subsequently, fibres were transferred to distilled water for 1 hour to remove excess buffer and solvent and then air dried.

**Table 1:** Experimental conditions investigated for the formation of wet spun gelatin fibres.

| Sample ID | Wet spinning solution (% w/v) | Coagulating bath |
|---|---|---|
| **Type A gelatin** | 15 | PEG 20% (w/v), 0.1 M PBS |
| **Gel-4VBC** | 15 | NaCl 20% (w/v) |
|  |  | PEG 20% (w/v), 0.1 M PBS |
| **Gel-MA** | 30 | EtOH |

*2.6 UV-curing of wet spun fibres*

Wet spun fibres were incubated in a 90 vol.% ethanol solution in distilled water supplemented with 1% (w/v) Irgacure 2925 for 30 min during UV exposure ($\lambda$: 365 nm, 8 mW·cm$^{-2}$) with a sample-UV lamp distance of approximately 10 cm. Irradiation intensities were measured with an IL1400A radiometer equipped with a broadband silicon detector (model SEL033), a 10× attenuation neutral density filter (model QNDS1), and a quartz diffuser (model W)

(International Light Technologies, USA). Fibres were then incubated in pure ethanol for 1 hour and air dried on a polyethylene surface.

### 2.7 Manufacture of nonwoven fabrics

Samples intended for fibroblast attachment were directly spun-laid into a nonwoven fabric by manually agitating the coagulation bath using a 2 cm wide implement placed eccentrically in the bath during rotation-assisted wet spinning. In this way, the fibre-forming wet spinning solution was displaced to enable merging of resulting fibres and nonwoven structure via drying-induced hydrogen bonding. PCL scaffolds were wet-spun for use as a positive control during cytotoxicity screening using a 15 wt.% PCL solution dissolved in acetone at 40 ºC and spun into ethanol, following a previous report (47).

### 2.8 Attenuated Total Resistance Fourier Transform Infrared (ATR-FTIR)

ATR-FTIR was performed to detect the presence of any PEG residue in wet spun fibres. Spectra were recorded on a Spectrum BX (Perkin-Elmer, USA) with a diamond ATR attachment at a resolution of 4.0 cm$^{-1}$, scanning intervals at 2.0 cm$^{-1}$. Scans were recorded from 600 to 4000 cm$^{-1}$ and 64 repetitions were averaged for each spectrum. A blank sample was measured to remove unwanted noise from atmospheric organic compounds.

### 2.9 Scanning Electron Microscopy

A SU8230 FESEM (Hitachi, Japan) was used to study longitudinal and cross-sectional fibre morphology. Samples were gold-coated with at a beam intensity of 10 kV in vacuum using a JFC-1200 fine sputter coater and imaged at ×20-1500 magnification (working distance 38-14 mm). ImageJ software was used to analyse images and quantify fibre diameter distribution.

### 2.10 Mechanical testing

Samples were conditioned at 20 °C 65% r.h. for 24 h. Fibres were mechanically tested using a Titan Universal Strength Tester (James Heal, UK) with a 10 mm gauge length and 10 mm·min$^{-1}$ extension rate following BS EN ISO 5079:1996 standards (n=10). Rubber pads were placed over the clamps to prevent breakage at the jaws. Any jaw breakages were discounted from results. Optical microscopy at ×4 magnification was used in conjunction with ImageJ analysis software to approximate the cross-sectional diameter of fibres to enable fibre stress calculations assuming a circular fibre shape.

## *2.11 Differential Scanning Calorimetry*

Thermal characterisation was performed using differential scanning calorimetry (DSC, Q20 V24.11 Build 124, TA Instruments, USA). Dry samples (11 mg) were hermetically sealed in aluminium pans and thermograms recorded with a nitrogen purge gas flow with a ramp of 0-200 $^{\circ}$C and incremental rate of 10 $^{\circ}$C·min$^{-1}$. OriginPro was used to plot DSC thermograms, whereby the denaturation temperature and enthalpy were quantified as the endothermic peak and integration of corresponding thermal transition area, respectively.

## *2.12 Fibre and nonwoven scaffold swelling*

Swelling of individual fibres was measured by imaging 1 cm length of fibres using optical microscopy before and after 24-hour incubation in PBS at 37ºC (n=4). Image analysis was used to calculate fibre diameters at a minimum of 10 points along the fibre length. These diameters were used to calculate the fibre cross-sectional area based on a cylindrical shape (A= π·r$^2$), so that dry- ($X_d$) and swollen- ($X_s$) state cross-sectional areas were obtained. The fibre swelling index (*SI*) was calculated as percent change in cross-sectional area following sample equilibration with water (Equation 3):

$$SI = 100 \times \frac{X_s - X_d}{X_d}$$  **Equation 3**

Other than fibre swelling, weight-based water uptake (*WU*) of nonwoven fabrics was quantified according to Equation 4:

$$WU = 100 \times \frac{m_s - m_d}{m_d}$$  **Equation 4**

whereby *$m_d$* and *$m_s$* are the sample weights before and after 24-hour incubation in PBS at 37ºC (n=6), as described previously (48).

*2.13 Cytotoxicity testing*

100 mg of dry nonwoven sample were individually placed in non-tissue culture treated 24-well plates and incubated overnight in a 70 vol.% ethanol solution in water. Retrieved samples were washed in sterile PBS three times under UV light, prior to cell seeding. L929 murine fibroblasts were grown in a Dulbecco's Modified Eagles Medium (DMEM) of 10 vol. % FBS and 1% (w/v) Penicillin Streptomycin and 1% (w/v) Glutamine. Cells were seeded (cell density: 8×10$^4$ cells in 100 µl cell culture medium) on each nonwoven scaffold and incubated for 8 hours to allow the cells to adhere to the scaffold. Cell-seeded nonwoven scaffolds were transferred to a newly sterilised 24-well plate and incubated for 7 days at 37°C. After incubation, the nonwoven scaffolds were washed with sterilised PBS (×3) before cell staining with Calcein AM and ethidium homodimer 1 (ThermoFisher Scientific, UK), according to manufacture instructions. The samples were then incubated for 40 min in the dark and imaged by fluorescence microscopy (DMI6000 B, Leica Microsystems, Germany). The cell viability/proliferation assay was assessed using an alamarBlue metabolic assay at days 1, 4, and 7. The alamarBlue working solution was measured at an excitation wavelength of 560nm and emission wavelength of 590nm using a fluorescence plate reader (Varioscan, ThermoScientific, USA).

*2.14 Statistical analysis*

OriginPro software was used for Normality tests, means comparisons using Tukey testing, and One-Way ANOVA providing normal distribution was observed. Data are presented as average ± standard error.

## 3. Results and Discussion

In the following, the design of photoactive and UV-cured gelatin fibres is presented and characterised from the molecular up to the macroscopic scale. The degree of gelatin functionalisation, the morphology, swelling and mechanical properties of wet spun fibres, as well as the manufacture and cellular tolerability of UV-cured gelatin nonwovens is reported.

Following synthesis of the photoactive gelatin building blocks, wet spun fibres and nonwovens were prepared using different coagulation conditions. The reacted wet spun gelatin product was collected and air dried prior to UV curing, as illustrated in Figure 1.

Functionalised gelatin samples are coded as "XXXXYY", whereby "XXXX" identifies the monomer, either MA or 4VBC. "YY" indicates the monomer/lysine molar ratio used during the functionalisation reaction (either 10 or 25). Sample nomenclature for the wet spun and UV cured fibre samples is as follows: '4VBC-PEG' indicates fibres whereby 4VBC-functionalised gelatin was dissolved in 17.4 mM acetic acid (15 wt.% gelatin) and wet spun into 0.1 M PBS supplemented with 20% w/v PEG. '4VBC-NaCl' indicates fibres prepared by dissolving 4VBC-functionalised gelatin (15 wt. %) in 17.4 mM acetic acid followed by wet spinning into an aqueous solution containing 20% w/v NaCl. 'MA-EtOH' identifies fibres prepared with 30 wt. % solution of MA-functionalised gelatin in acetic acid and wet spun into ethanol at -20ºC; 'Native-PEG' is the fibre control made of 15 wt. % native gelatin in 17.4mmol acetic acid and wet spun into PEG-supplemented PBS bath; 'Native-NaCl' is the fibre control prepared by wet spinning the native gelatin solution into the aqueous solution containing 20 % w/v NaCl.

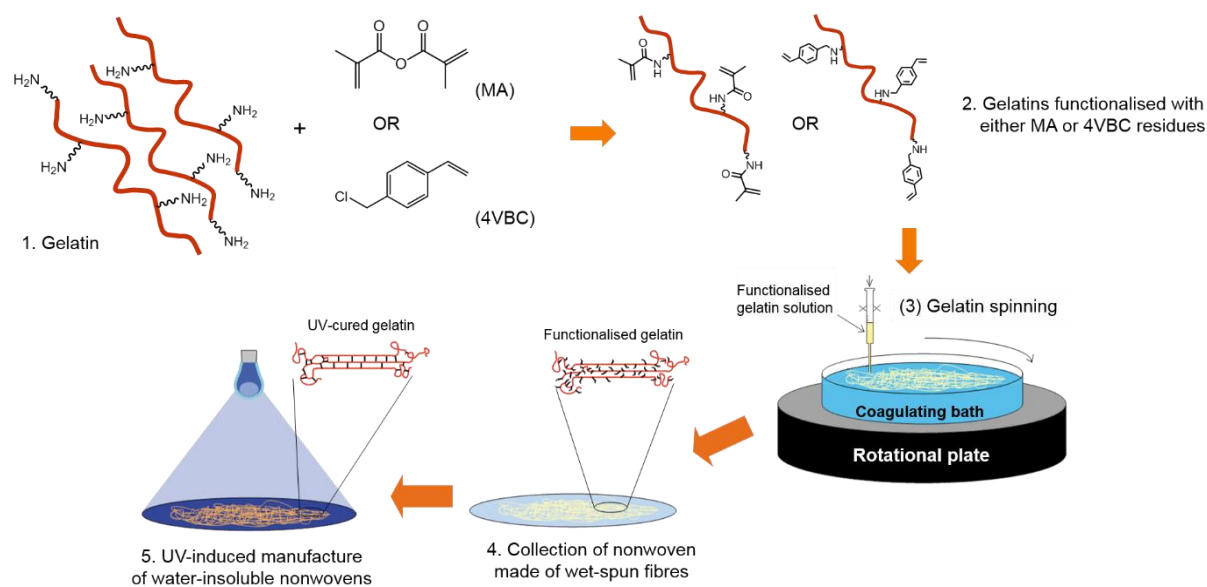

**Figure 1.** Rotation-assisted wet spinning of UV-cured gelatin fibres and nonwovens. Gelatin is reacted with either MA or 4VBC (1) to achieve a photoactive fibre-forming building block (2). The functionalised gelatin product is dissolved in aqueous environment and wet spun into a rotating coagulating bath (3). Either the individual fibres or nonwovens are withdrawn from the bath and incubated in a photoinitiator-supplemented ethanol solution (4). UV curing leads to water-insoluble fibres and nonwovens made of a covalently-crosslinked gelatin network (5).

### *3.1 Synthesis of functionalised gelatin*

TNBS colorimetric assay was employed in reacted gelatin products to indirectly assess the derivatisation of lysine amino groups of gelatin with either 4-vinylbenzyl or methacrylamide adducts. Reaction products indicated that both 4VBC- and MA-mediated functionalisation were achieved, with degree of functionalisation (*F*) equalling to 66±4 and 97±2 mol.%, respectively (Table 2). The consumption of amino groups measured by TNBS was found to increase when gelatin was reacted with MA rather than 4VBC, despite the same monomer/Lys ratio being used during the reaction. In an effort to achieve building blocks with equivalent molar content of 4-vinylbenzyl and methacrylamide adducts, both MA/Lys molar ratio (25→10 mol.%) and reaction time (5→3 hours) were reduced during the functionalisation reaction, although insignificant variations (p=0.99) in values of *F* were observed with reacted products ($F_{MA}$= 99±0 mol.%). These results are in line with previous reports (34, 49) and reflect the higher reactivity of MA towards lysine-induced nucleophilic substitution compared to 4VBC.

**Table 2.** Degree of functionalisation (*F*) determined via TNBS assay on gelatin products reacted with either 4VBC or MA at varied monomer/Lys molar ratio. [a-c]: $p < 0.05$.

| Sample ID | *F* /mol.% |
|---|---|
| **MA10** | 97±2 [a] |
| **MA25** | 99±0 [b] |
| **4VBC10** | 47±4 [b] [c] |
| **4VBC25** | 66±4 [a] [c] |

Interestingly, the *F* values observed with MA- and 4VBC-reacted gelatin were higher than those previously reported with 4VBC-reacted type I rat tail collagen, when the same reaction conditions were applied [26]. Despite collagen and gelatin sharing the same primary protein structure, the reversible helix-to-coil thermal transition of gelatin allows for increased chemical accessibility in the selected reaction conditions, with regards to the steric hindrance effects expected when the same reaction is carried out with collagen triple helices (50).

In comparison to commercially-available Gel-MA products ($F_{MA}$= 40-80%), the values of $F_{MA}$ measured on in-house-reacted gelatin products proved to be increased. This finding is potentially due to an increased molar content of either MA or TEA, or to an increased reaction time being used for the synthesis in-house of Gel-MA product, as reported previously (51). This suggests that the developed reaction conditions used herein may enable increased tunability of macroscopic properties and fibre-forming capability in resulting products. In light of the presented results, samples MA10 and 4VBC25 were selected for the preparation of wet-spinning solutions and respective fibres, as reported in the following sections.

## *3.2 Characterisation of wet-spinning solutions*

The viscosity of the wet spinning solutions was measured to assess the levels of shear force being exerted upon the spinning solutions during extrusion, aiming to draw relationships between solution characteristics and fibre dimensions and morphology.

From Figure 2, it is apparent that all gelatin solutions behave as a non-Newtonian fluid, whereby an increase in shear rate is corelated with a decrease in shear viscosity ($\eta$), in line with previous studies (49, 52). When the shear rate is increased, gelatin chains disentangle since molecular interactions are attenuated, and the corresponding solution viscosity is reduced. As expected, an increase in Gel-MA concentration resulted in an increase in viscosity, due to the increased chain entanglements and intermolecular interactions (49, 52).

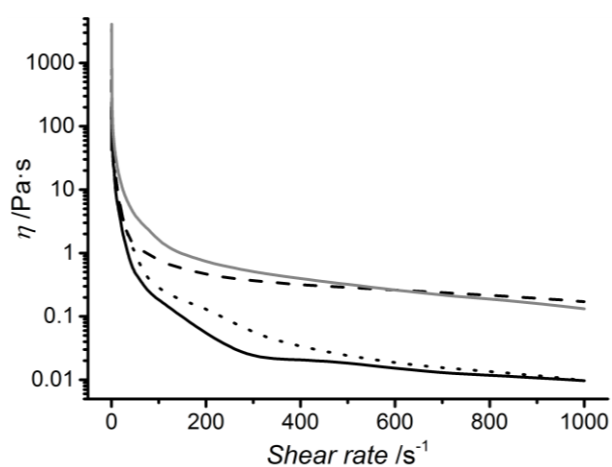

**Figure 2.** Viscosity curves of wet spinning solutions prepared in 17.4 mM acetic acid solution with varying gelatin building blocks. (—): Native gelatin (15 wt. %). (⋯): Gel-4VBC (15 wt. %). (– – –): Gel-MA (15 wt. %). (—): Gel-MA (30 wt. %).

With respect to the native polypeptide, functionalisation of gelatin was found to lead to an increased viscosity when solutions with the same gelatin concentration were measured. This observation is likely due to an increase in lysine side chain length in functionalised gelatin, causing more entanglement and interaction between polypeptide chains. Similar trends were also observed when considering solutions of Gel-4VBC, in light of the presence of 4VBC aromatic rings and pi-pi interactions between lysine residues (34) (53).

Other than Gel-Native, the wet spinning solutions of Gel-MA were considerably more viscous than the ones containing Gel-4VBC, when the same concentration of polypeptide was considered. This is likely to reflect the increased level of functionalisation in Gel-MA with respect to Gel-4VBC.

Although the original intention was to compare wet spun fibre produced with wet spinning solutions of similar polymer concentrations, it was not possible to continuously wet spin fibres from solutions of Gel-MA with 15 wt.% concentration (as in the case of solutions of Gel-4VBC) due to dispersion of the wet spinning solution in the coagulating bath. Therefore a solution of 30 wt.% Gel-MA was selected, in agreement with previous literature (19).

Such an increase in polypeptide concentration, solution viscosity and internal shear stress during wet spinning was likely to result in reduced alignment of the polypeptide chains in respective wet spun fibres (28, 30). These considerations were experimentally supported by the fact that needle blockages as well as difficulties in achieving continuous fibre formation were encountered during wet spinning of solutions with increased polypeptide concentration.

### *3.3 Chemical characterisation of wet spun gelatin fibres*

Dried wet spun fibres were analysed via ATR-FTIR spectroscopy to detect the presence of any chemical residues, e.g. PEG. Figure 3A compares the spectra of fibres spun into PEG-supplemented PBS solution with the ones of gelatin and PEG raw materials. The gelatin spectra show a major peak at 3600-2700 $cm^{-1}$ (Amide A) as well as peaks at roughly 1630 (Amide I) and 1550 $cm^{-1}$ (Amide II). An additional peak at 1230 $cm^{-1}$ (Amide III) was also observed in accordance with previous literature (54-57). Other than the gelatin peaks, the ATR-FTIR spectra was enlarged close to the fingerprint area to further elucidate the presence of PEG in the fibres (Figure 4B). The spectra showed clear disparity between the gelatin fibres spun in to PEG-supplemented buffer and the native gelatin raw material. Clear peaks related to ether bonds (at 1100 $cm^{-1}$) were observed in the spectra of both PEG raw material and gelatin fibres spun into the PEG-supplemented bath. Although this peak was slightly shifted towards reduced wavenumbers due to the interference of gelatin amide bonds, the result confirms the presence of PEG in the wet spun fibres, as reported previously (58, 59).

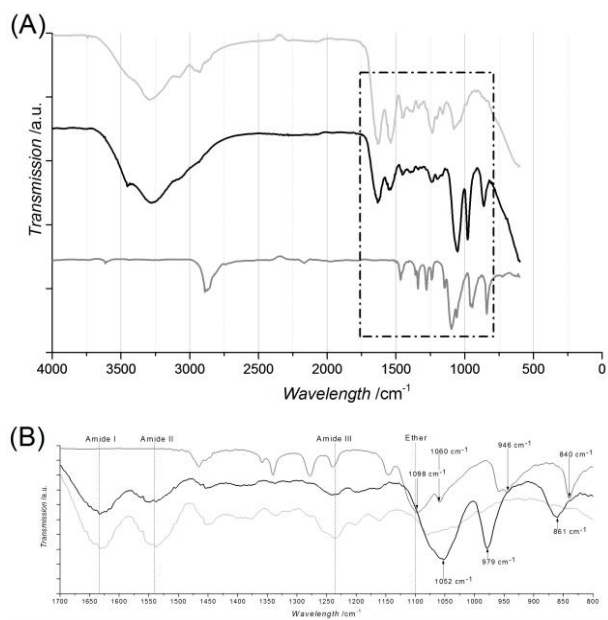

**Figure 3.** ATR-FTIR spectra of fibres made of native gelatin wet spun in PEG-supplemented PBS coagulation bath (—) as well as PEG (—) and gelatin (—) raw materials. Both full (A) and zoomed-in (B) spectra are reported.

### 3.4 Fibre morphology

Other than the chemical composition, collected wet spun fibres were inspected by SEM to gather qualitative and quantitative information on the fibre morphology. As expected, the employment of a static coagulating bath during wet spinning process led to decreased fibre uniformity (Figure 4A), as compared to samples wet spun into rotating coagulation bath (Figure 4B). Although low (~ 1.6 rpm), the rotation of the coagulating bath was expected to induce a degree of fibre drawing and increased molecular chain alignment of the fibre-forming polymer, resulting in fibres with increased surface homogeneity, thermal stability and tensile modulus (60, 61). In contrast to statically wet spun fibres (Figure 4A), the rotary wet spinning process clearly enabled the formation of fibres with smooth surface morphology and few cross sectional cavities (Figure 4 B-C). Although surface roughness may aid cellular attachment and alignment, they are known to induce variation in fibre strength, and undesirable variation in mechanical properties (19, 62-64).

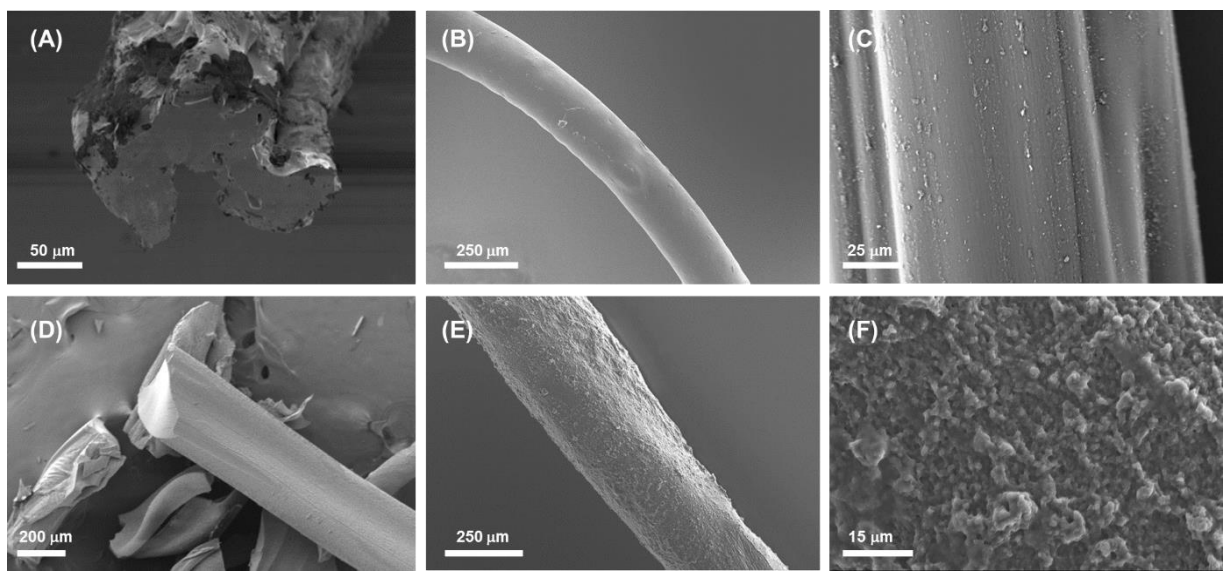

**Figure 4.** SEM images of wet spun gelatin fibres. (A): 4VBC-PEG. (B-C): 4VBC-PEG. (D): MA-EtOH. (E-F): 4VBC-NaCl. Image (A) refers to a sample wet-spun in a static coagulating bath. All other images refer to samples wet spun in a rotating coagulating bath.

Other than the effect of the coagulating bath rotation, the selection of specific non-solvents for the wet spinning gelatin solutions also proved to markedly affect resultant fibre morphology. Although fibre diameters were comparable, fibres (*d*: 336 ± 5 μm) spun in the sodium chloride-supplemented coagulating bath appeared less regular and more opaque than fibres (*d*: 370 ± 4 μm) spun using PEG-supplemented PBS (Figure S1, Supplementary Information), suggesting differences in crystallinity between the two groups. With respect to the salt-supplemented solution, the increased translucency observed in the latter samples could be due to a PEG-induced reduction in crystallinity caused by the long PEG chains ($M_n$: 8000 g·mol$^{-1}$) intertwining within the gelatin polypeptides. In light of the well-known miscibility with gelatin solutions, PEG chains could impede the spinning-induced alignment of gelatin chains, so that the formation of amorphous materials with increased light transmission capability could be explained (Figure S1, Supporting Information).

In comparison to fibres made of Gel-4VBC, MA-EtOH fibres proved to display slightly decreased mean diameter (*d*: 322 ± 14 μm) and smooth surface morphologies, although they were found to be extremely brittle, as evidenced by top surface SEM (Figure 5 D-E).

The large diameters of the fibres can be explained considering that a spinneret needle with an 800 µm internal diameter was employed during wet spinning. It is expected that decreased fibre diameters can be obtained via fibre drawing during or following wet-spinning.

### *3.5 Fibre swelling and nonwoven liquid uptake*

Following characterisation of fibre morphology, the attention moved to the quantification of fibre properties in a hydrated near-physiological environment. Wet-spun fibres were UV-cured and characterised with regards to their liquid uptake capability. This is relevant for applications in medical devices such as wound dressings, where production of wound exudate can reach 4000-12000 g·m$^{-2}$·day$^{-1}$ (65, 66)), and for tissue scaffolds, where nutrient and waste exchange is mediated by the physiological medium (67).

The liquid uptake capability, hereby measured in terms of gravimetric and dimensional change, is also affected by the degree of covalent crosslinking achieved during fibre UV curing as well as molecular alignment and degree of crystallinity.

Fibre swelling and water uptake were measured for individual fibres and for nonwoven webs, respectively. For nonwovens, the liquid uptake also depends on liquid sorption of free liquid between the fibres, which is governed by the architecture of the structure. This leads to marked differences in water uptake as indicated Figure 5 A and B.

Photoinduced crosslinking was confirmed by investigating the macroscopic properties of UV-cured as well as wet spun Gel-4VBC fibres in near-physiologic conditions. UV-cured fibres proved to display significant swelling following 37 ºC incubation in aqueous environment (Figure 5 A), In contrast to the prompt dissolution observed in the case of the wet spun, non-UV-cured fibres. This observation therefore supports the fact that selected gelatin functionalisation and UV-curing route was key to enabling the synthesis of a covalent gelatin network at the molecular scale, and the manufacture of water-insoluble, mechanically-competent wet spun fibres at the microscale. This confirmation was further obtained at room temperature, whereby significantly higher swelling index was measured in wet-spun, non-UV-crosslinked fibres 4VBC-PEG (SI= 234 ± 43 a.%) and 4VBC-NaCl (SI= 153 ± 28 a.%, p= 0.04), in contrast to the UV-cured fibres 4VBC-PEG (SI= 184 ± 12 a.%, p= 0.046) and 4VBC-NaCl (SI= 71 ± 18 a.%, p= 0.04).

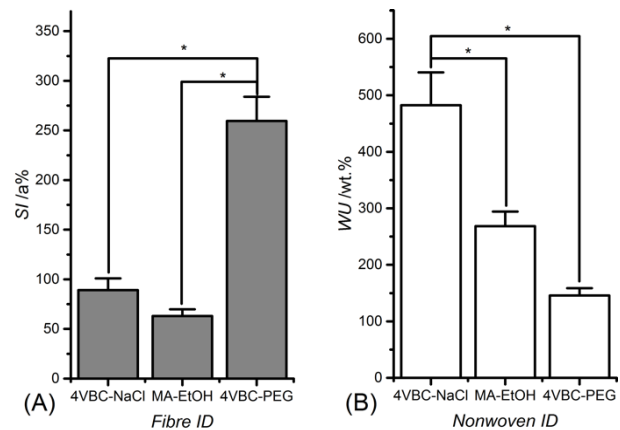

**Figure 5.** Swelling index (*SI*, A) of individual UV-cured wet spun fibres as well as water uptake (*WU*, B) of UV-cured nonwovens, following 24-hour incubation in PBS. (*: p<0.05, n=4).

Individual 4VBC-PEG fibres showed a significantly higher (p=0) level of swelling index (*SI*= 260±24 a%) than fibres 4VBC-NaCl (*SI*= 89±12 a% p=0) and MA-EtOH (*SI*= 64±7 a%). This is likely due to the hygroscopic nature of PEG present within the fibres, promoting an amorphous fibre morphology, allowing for liquid to move into the regions between the polypeptide chains. Other than the presence of PEG in 4VBC-PEG fibres, the reduced liquid uptake capability of MA-EtOH fibres is also in agreement with the increased degree of functionalisation found in MA- with respect to the 4VBC-reacted gelatin fibre building block.

With regards to the swelling trends observed with fibres, an almost inverse relationship was observed when quantifying the water uptake of corresponding gelatin nonwovens. Nonwoven samples made of 4VBC-NaCl fibres displayed the highest water uptake (*WU*: 482±58 wt. %), which was significantly higher than the one measured in nonwovens of either 4VBC-PEG (*WU*: 146±13 wt. %, p=4×10$^{-5}$) or MA-EtOH (*WU*: 268±26 wt. %, p=0.003) fibres. Given that the SEM images of respective nonwovens appeared similar among the different sample groups (Figure 8, A-D); the most likely explanation for the above observation is that the water-induced swelling of the individual fibres leads to reduced porosity in the corresponding water-equilibrated nonwoven. Consequently, decreased water content is expected within the pores of the nonwoven structure, when fibres with increased swelling capability are employed.

### 3.6 Thermal characterisation

Thermal properties of native gelatin as previously reported in the literature indicate a glass transition temperature of 80-90°C and a triple helix-related denaturation temperature of 110-115°C (1). In the present study, no defined glass transition was detected in the DSC thermograms of the UV-cured fibres (Figure 6), which may be due to the presence of free water residues in the tested samples (2).

UV-cured 4VBC-PEG fibres were associated with the highest denaturation temperature ($T_m$ =108 °C), whilst a denaturation enthalpy of 106 J·g$^{-1}$ was measured (Table 2). In comparison, UV-cured 4VBC-NaCl fibres displayed the lowest denaturation temperature ($T_m$ = 84°C), and a slightly higher enthalpy (ΔH= 125 J·g$^{-1}$). The denaturation enthalpy indicates the heat energy required to break the crystalline-like triple helical junctions within the gelatin material (3). During the wet spinning of 4VBC-PEG fibres, PEG is known to act as a "reptating agent" during fibre formation (6), intertwining with the fibre-forming polypeptide chains and leading to an increased melting point. In this instance, the presence of PEG within the polypeptide chains is expected to inhibit the renaturation of gelatin molecules into collagen-like triple helices, leading to decreased fibre denaturation enthalpy. This molecular organisation mechanism is likely missing in the case of fibres 4VBC-NaCl wet spun in PEG-free buffers. Here, the absence of PEG in the coagulating buffer enables chain alignment and renaturation of collagen-like triple helix to occur unperturbed (5), so that an increased fibre denaturation enthalpy can be measured with respect to fibres 4VBC-PEG. This mechanism is supported by both the observed morphology, swelling, and mechanical properties of the fibres.

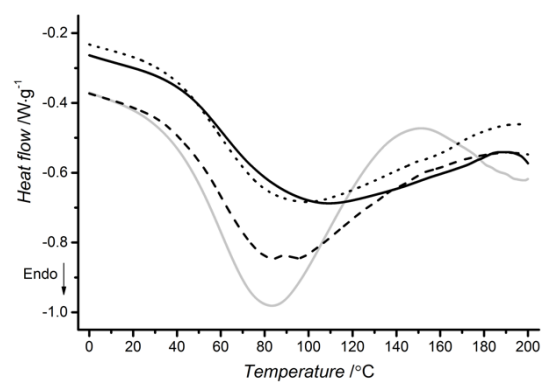

**Figure 6.** DSC curves of UV-cured wet spun fibres MA-EtOH (– – –), 4VBC-NaCl (—) and 4VBC-PEG (—) as well as fibre control Native-PEG (···).

## 3.7 Tensile properties

Together with the evaluation of liquid uptake and thermal properties, tensile measurements were also carried out on individual UV-cured and native gelatin fibres.

The ultimate tensile strength (*UTS*) was significantly higher in the UV-cured 4VBC-PEG fibres (*UTS*: 74±3 MPa) with respect to samples MA-EtOH (p=0), 4VBC-NaCl (p=8×$10^{-9}$) and Native-PEG (p=9×$10^{-7}$) (Figure 7). Remarkably, the tensile strength of water-equilibrated 4VBC-PEG samples was also sufficiently high to permit knotted and twisted fibre conformations to be generated by hand (Figure S1, Supplementary Information), providing further evidence of the mechanical stability of these fibres and their potential applicability as materials for medical device manufacture.

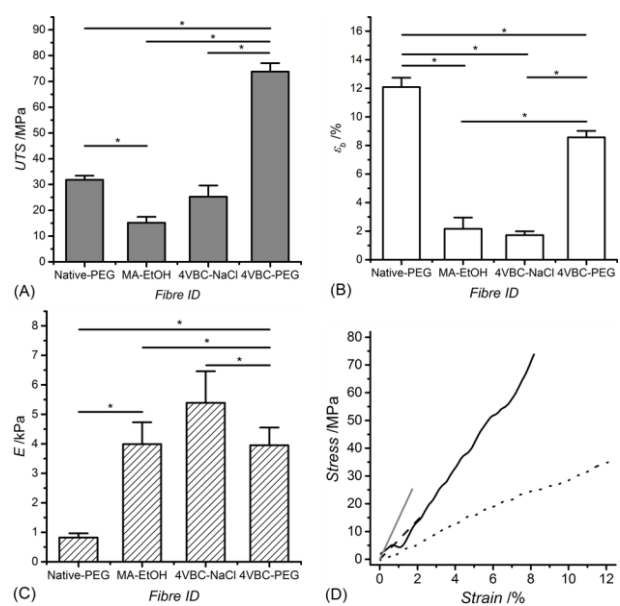

**Figure 7.** Tensile properties of individual wet-spun fibres made of either native or UV-cured gelatin. (A): Ultimate tensile stress (*UTS*); (B): extension at break ($\varepsilon_b$); (C): Tensile modulus (*E*); (D): Averaged stress-strain curves of UV-cured wet spun fibres MA-EtOH (‒ ‒ ‒), 4VBC-NaCl (—), and 4VBC-PEG (—) as well as fibre control Native-PEG (⋯).*: p<0.05.

The increased values of tensile strength and elastic modulus measured with fibres made of 4VBC-PEG and 4VBC-NaCl with regards to fibre MA-EtOH, reflect molecular scale considerations regarding the UV-induced 4VBC-based crosslinking segment as well as the nature of the wet spinning coagulating bath. UV-curing of 4VBC-functionalised gelatin chains leads to the formation of an aromatic crosslinking segment, with increased molecular rigidity with respect to the methacrylate segment obtained via UV-curing of MA-functionalised gelatin chains. Besides the inherent molecular rigidity of the UV-induced crosslinking segment, the size of the photoactive moiety covalently-coupled to the gelatin backbone is also expected to impact on the yield of the free-radical crosslinking reaction. Methacrylamide residues in MA-functionalised gelatin are indeed shorter than the 4-vinylbenzyl moieties present along the 4VBC-functionalised gelatin backbone, so that so that steric effects potentially hindering the free-radical crosslinking reaction are likely. These observations may count for the trends observed in tensile strength and tensile modulus with respect to the UV-cured fibres MA-EtOH and 4VBC-NaCl (Table 3).

Other than that, it is likely that the plasticising effect of PEG coagulant contributed to the increase in strength of the UV-cured 4VBC-PEG fibres. Not only does PEG provide a more viscous coagulant in which diffusional exchange during the wet-spinning process can occur more slowly preventing the formation of voids within the fibres (20), but the confirmed presence of PEG in the fibres is also expected to increase the tensile strength via hydrogen bonding between the polymer and gelatin hydroxyl groups (42, 68). The PEG polymer chains have previously been reported to act as a reptating agent, entangling within the chain structure and sliding parallel to gelatin chains when withstanding tensile forces (42, 69).

In contrast to PEG-containing fibres, wet spinning in the NaCl-supplemented coagulating bath resulted in extremely weak and brittle fibres. In the case of MA-functionalised gelatin, resulting wet spun fibres were too fragile to handle and so no mechanical data could be obtained. Likewise, 4VBC-NaCl samples were very brittle and, despite their increased Young's modulus ($E$: 5000 ± 1000 kPa), they could be uniaxially extended only up to about 2% of their initial length prior to fibre break.

**Table 3.** Thermo-mechanical properties of gelatin fibres. UTS: Ultimate tensile strength; $E$: Young's modulus, $T_m$: melting temperature; $\Delta H$: melting enthalpy

| Sample ID | UTS /MPa | Strain /% | E /kPa | $T_d$ /ºC | $\Delta H$ /W·g$^{-1}$ |
|---|---|---|---|---|---|
| **4VBC-NaCl** | 25±4 | 1.7±0.3 | 5000±1000 | 84 | 125 |
| **4VBC-PEG** | 74±3 | 8.6±0.5 | 4000±1000 | 108 | 106 |
| **Native-PEG** | 9±1 | 12.1±0.7 | 300±40 | 99 | 101 |
| **MA-EtOH** | 15±2 | 2.2±0.8 | 4000±1000 | 89 | 102 |

The observed reduction in extension at break in samples 4VBC-NaCl with respect to 4VBC-PEG samples is in agreement with the fact that no gelatin plasticiser, i.e. PEG, was present. This tensile data also suggest that the coagulating bath was not sufficiently viscous to support the diffusional exchange occurring during fibre formation, therefore resulting in the rapid diffusion of solvent away from the fibre-forming gelatin chains and the creation of weak fibres with irregular internal structure. This is reflected by the morphological characterisation (Figure 5E and F), which revealed fibres with a rough surface and low uniformity. The presence of PEG-induced plasticising effect in fibres with increased elongation at break agrees with previously discussed fibre thermal properties, whereby a corresponding decrease in denaturation enthalpy was recorded (Table). The decreased denaturation enthalpy measured in both samples Native-PEG and 4VBC-PEG correlates with a decrease in fibre crystallinity, whereby resulting amorphous regions mediate fibre elongation during the tensile test. During stretching, fibre elongation is indeed mainly attributed to the straightening of amorphous regions, whilst the presence of chemical bonds fixes gelatin chains in a molecular network (70).

Other than the UV-cured fibres, it was interesting to note that the native gelatin control fibres were significantly more extensible with respect to samples MA-EtOH ($p=8\times10^{-9}$), 4VBC-NaCl ($p=0$), and 4VBC-PEG ($p=2\times10^{-4}$). This was found to correlate with the fact that no covalent crosslinking was introduced between the fibre-forming gelatin chains, therefore allowing for parallel chain slippage to occur during uniaxial tension.

### 3.8 Morphology and cytotoxicity of gelatin nonwovens

To determine the feasibility of forming fabrics directly during the wet spinning process, nonwoven webs were spunlaid from wet spun gelatin fibres, using a rotary deposition system and slight agitation of the coagulating liquid. Agitation of the coagulation bath allowed displacement and crossing of proximal fibres to produce a self-supporting fibrous network that could be manually manipulated after drying. The fibre morphology in the nonwoven assembly varied considerably depending upon the fibre building block and the selection of the coagulating liquid (Figure 8). Unlike the case of the individual wet spun fibres, respective nonwovens 4VBC-NaCl exhibited irregular fibre morphology (Figure 8 A-B).

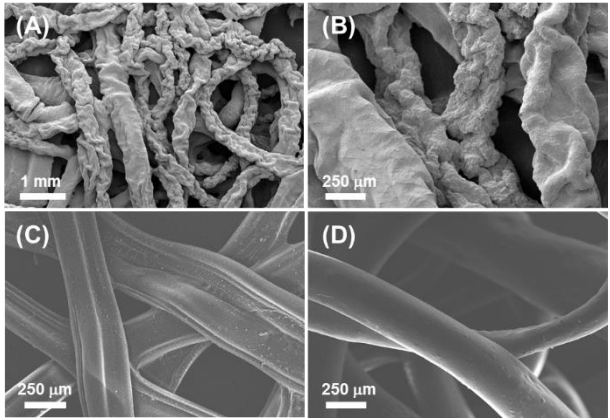

**Figure 8.** SEM micrographs of UV-cured gelatin nonwovens prepared during wet-spinning in one step. (A-B): 4VBC-NaCl; (C): MA-EtOH; (D): 4VBC-PEG.

This observation is likely due to merging of fibres during wet spinning and prior to drying, leading to surface buckling and collapse and explaining the morphology variation between the nonwoven and single fibre(71). MA-EtOH fibres (Figure 8 C) exhibited flattened or ribbon-like morphologies, again suggesting rapid diffusion of the dope solvent from the fibre-forming gelatin solution. By contrast, 4VBC-PEG fibres (Figure 8 D) were characterised by a smooth and rounded morphology with negligible fibre merging, indicating coagulating bath conditions consistent with enhanced diffusional exchange in light of the increased viscosity of the coagulating bath.

The cellular tolerability of the nonwoven structures containing gelatin was assessed with L929 murine fibroblasts following 7-day cell culture via both an alamarBlue assay and Live/Dead cell staining, whereby a PCL nonwoven control was also tested (Figure 9). All gelatin nonwovens showed significantly higher ability to support cell survival than the PCL control, following both 4- and 7-day cell culture. These results confirm that the increased cellular tolerability of gelatin with respect to PCL is maintained even though chemical functionalisation, fibre spinning and UV-induced network formation were applied to the native polypeptide (47). Within the two gelatin groups, a significant difference in metabolic activity was observed in cells cultured on fibres made of 4VBC-functionalised gelatin when compared to cells cultured on fibres made of MA-functionalised, following 7-day cell culture. This could be due to multiple factors such as the increased degree of functionalisation and corresponding crosslink density measured in methacrylated gelatin fibre, which may mask the cell-binding sites present in native gelatin.

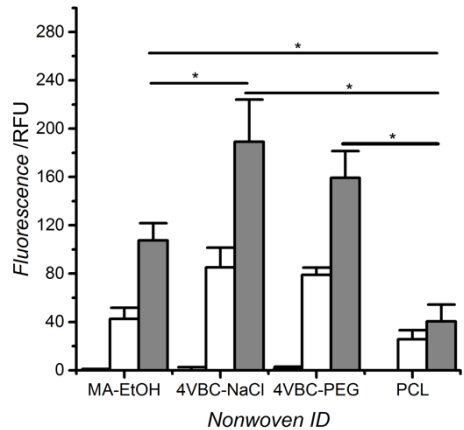

**Figure 9.** alamarBlue assay following L929 cell culture on both UV-cured gelatin nonwovens and PCL controls at day 1 (black columns), 4 (white columns), 7 (grey columns). *: p< 0.05.

Within the 4VBC-functionalised gelatin groups, an insignificant decrease in metabolic activity was recorded in cells following 7-day culture on sample 4VBC-PEG compared to cells cultured on sample 4VBC-NaCl, suggesting that the presence of PEG within the fibre did not induce any detectable change in cell response. Overall, confocal images from the Live/Dead staining supported previous results, whereby substantial cellular viability was observed within the nonwoven gelatin fibres (Figure 10 A-D),

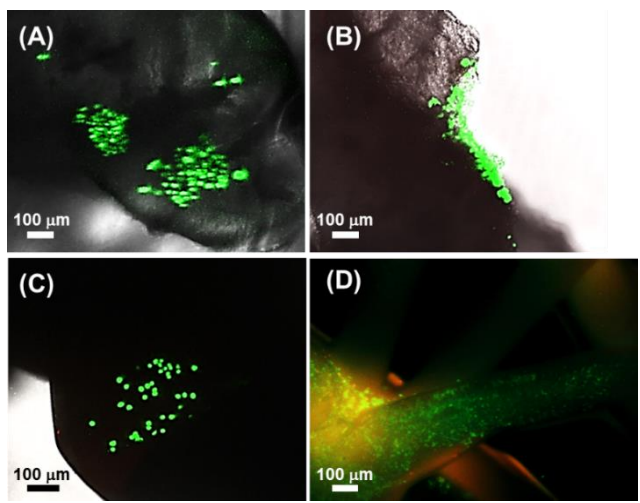

**Figure 10.** (A-C): Confocal images obtained following Live/Dead staining of L929 fibroblasts cultured on to UV-cured gelatin nonwovens for 7 days. (A): 4VBC-PEG; (B): 4VBC-NaCl; (C): MA-EtOH. (D): Fluorescence microscopy image of UV-cured gelatin nonwoven MA-EtOH following 7-day culture with L929 fibroblasts and Live/Dead staining.

## 4. Conclusions

Both 4VBC- and MA-functionalised gelatin precursors were successfully wet spun and UV-cured in both the fibre and nonwoven state, demonstrating the formation of mechanically-competent building materials for use in medical devices. The type of photoactive monomer proved to impact on the degree of gelatin functionalisation, so that solubility and spinnability of the resulting product were affected. Solutions of 4VBC-functionalised gelatin (15 wt. %) were readily spun at room temperature into a non-volatile aqueous coagulating bath supplemented with either NaCl or PEG. On the other hand, wet spinning solutions of MA-functionalised gelatin required a two-fold increase in gelatin concentration and a -20 ºC ethanol-based coagulating bath for fibres to be successfully formed. Regardless of the coagulating parameters, 4VBC-based wet spun fibres compared favourably with respect to MA-based fibres in terms of biocompatibility when tested with murine fibroblasts for up to 7 days. These data therefore indicate that the photoactive gelatin precursor functionalised with 4-vinylbenzyl residues enables enhanced fibre spinnability and cellular tolerability, which is key to enable further development for use in medical device manufacture spinning on a full scale set up, using a spinneret, and whilst fully controlling parameters such as draw ratio. Both the spinning process and gelatin building block developed in this study present wide applicability in bioprinting, whereby a computer-controlled extrusion device could be employed to deposit and UV-cure layers of gelatin fibres aiming to achieve precise 3D architecture of fibres (72-74). Variation in wet spinning coagulating bath conditions was shown to produce a significant difference in the morphology and mechanical properties of the fibres. 4VBC-functionalised gelatin solutions spun into a PEG-supplemented PBS buffer generated fibres with increased ultimate tensile strength ($UTS$= 74±3MPa) and strain at break ($\varepsilon_b$= 8.6±0.5 %) compared to fibres spun into a sodium chloride-supplemented aqueous solution ($UTS$= 25±4 MPa; $\varepsilon_b$= 1.7±0.3 %). As confirmed by ATR-FTIR, this variation of mechanical properties was likely due to the uptake of PEG molecules into the forming wet spun fibres, acting as a reptating agent within the polypeptide chains of the fibre. This PEG-induced effect also explains the homogenous morphology observed in fibres spun against the PEG-supplemented buffer with respect to the case of PEG-free spinning.

**Acknowledgements**

For their training and assistance, the authors would like to thank Jackie Hudson and Sarah Myers of the Department of Oral Biology (School of Dentistry), and Jianguo Qu and Dr. Tim Smith from the School of Design. Financial support from the Clothworkers Centre for Textile Materials Innovation for Healthcare is also gratefully acknowledged.

**Conflicts of Interest:** The authors declare no conflict of interest.

**Data Availability**

The raw/processed data required to reproduce these findings cannot be shared at this time due to technical or time limitations.

**Supplemental Materials**

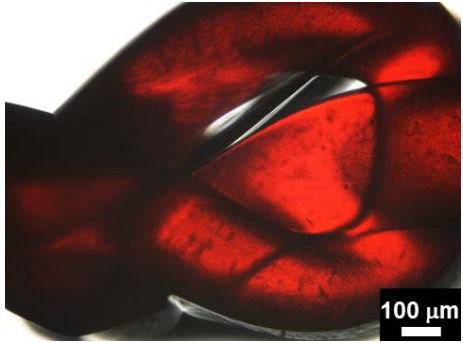

**Figure S1.** Optical microscopy image of a knotted sample of UV-cured gelatin fibre 4VBC-PEG following PBS equilibration.

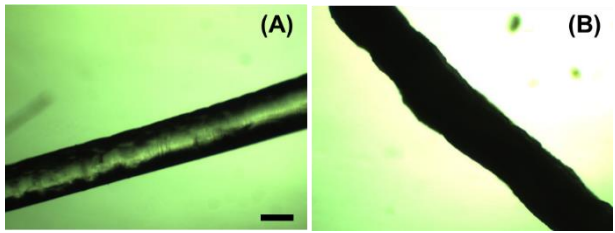

**Figure S2.** Optical microscopy images of UV-cured wet-spun fibres made of 4VBC-functionalised gelatin following 24-hour conditioning at 20 °C and 65% r.h. (A): 4VBC-PEG; (B): 4VBC-NaCl. Scale bar (200 μm) applies to both images.